\documentclass[letterpaper,twocolumn,fleqn]{article}

\usepackage{ist}
\usepackage{multirow}
\usepackage{array}
\usepackage{url}

\title{Characterization of Search Engine Caches}

\author{Frank McCown, Michael L. Nelson, Old Dominion University; Norfolk, Virginia/USA}

\date{} 

\begin{document}

\maketitle

\thispagestyle{empty} 


\begin{abstract}

Search engines provide cached copies of indexed content so users will have something to ``click on'' if the remote resource is temporarily or permanently unavailable.  Depending on their proprietary caching strategies, search engines will purge their indexes and caches of resources that
exceed a threshold of unavailability.  Although search engine caches are provided only as an aid to the interactive user, we are interested in building reliable preservation services from the aggregate of these limited caching services.  But first, we must understand the contents of search engine caches.  In this paper, we have examined the cached contents of Ask, Google, MSN and Yahoo to profile such things as overlap between index and cache, size, MIME type and ``staleness'' of the cached resources.  We also examined the overlap of the various caches with the holdings of the Internet Archive\footnote{Permission to make digital or hard copies of part or all of this work for personal or classroom use is granted without fee
provided that copies are not made or distributed for profit or commercial advantage and that copies bear this notice and the
full citation on the first page. Copyrights for components of this work owned by others than IS\&T must be honored.
Abstracting with credit is permitted. To copy otherwise, to republish, to post on servers or to redistribute to lists, requires
prior specific permission and/or a fee.\linebreak \emph{IS\&T Archiving 2007}, May 21--24, 2007, Arlington, Virginia, USA.}.


\end{abstract}

\section{Introduction}
\label{sec:intro}

To provide resiliency against transient errors of indexed web pages,
most search engines (SEs) provide links to cached versions of many of the
resources they have indexed.  Unlike the Internet Archive (IA), these SE
caches do not represent an institutional commitment to preservation.
Rather, they are intended to provide a link to the most recently
crawled version of the resource if the current resource is unavailable.
Sometimes the caches are not of the original resource, but the resource
migrated to new a format (e.g., PDF to HTML).

At Old Dominion University, we are engaged in a number of research
projects that utilize SE caches as the building blocks for digital
preservation services.  This includes the ``lazy preservation''
project \cite{McCown2006:Lazy}, which uses the IA and SE caches as
a preservation strategy and the ``just-in-time preservation'' project
\cite{Harrison:Just-In-Time}, which uses the IA and SE caches to generate
lexical signatures of missing resources to aid in the discovery of new
or similar versions of the missing resource.  Even though the SE caches
are not ``deep'' like the IA, they are very broad and are quite useful
in complimenting the IA's holdings.  We know that SEs do not always
immediately purge their caches if the original resources are unavailable.
For example, in a previous experiment we observed in Google's cache resources
that had been unavailable from the original source, and even missing from their cache,
for more than two months \cite{smith:observed}.

Since much of our research relies on SE caches, we have undertaken what
we believe to be the first quantitative analysis of SE cache behaviors
and contents.  We examined the caches of Ask, Google, MSN and Yahoo
by issuing dictionary-based queries to the SEs and characterizing the
caches of the returned resources.  In particular, we measured mean file
size, file MIME type, age of the cached resource, cache errors (i.e.,
the resource is declared as cached but not retrievable) and ``cache-only''
availability (i.e., the original resource is unavailable as indexed and
the cached version is available).  We also measured the overlap of the SE
caches with the IA and computed the ``staleness'' of the cached resources.
Finally, we uncovered a number of cases where the SEs had cached
what they arguably should not have (e.g., resources with a ``Cache-Control:
Private'' header).

\section{Background and Related Work}

\begin{figure}[!hb]
  \scalebox{0.51}{\includegraphics[clip=false]{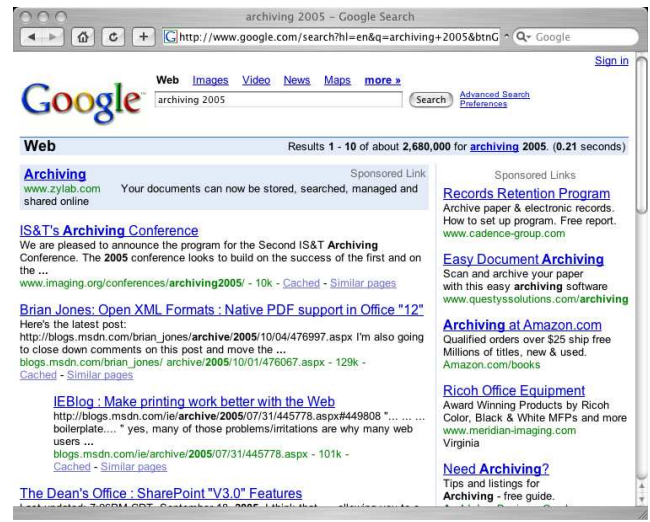}}
  \caption{Google search for ``Archiving 2005''.}
  \label{fig:google-search-2005}
\end{figure}

\begin{figure}[!hb]
  \scalebox{0.51}{\includegraphics[clip=false]{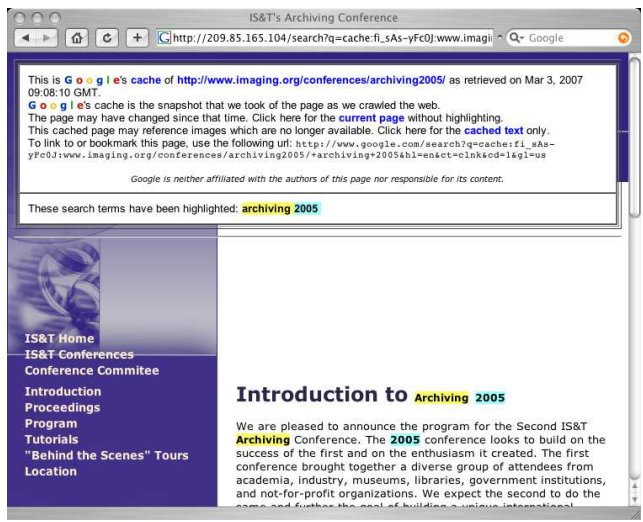}}
  \caption{Google's cached copy with datestamp of March 3, 2007.}
  \label{fig:google-cache-2005}
\end{figure}

Figure \ref{fig:google-search-2005} shows the results of searching Google
for ``Archiving 2005''.  The first result is what we expect, and next
to the URL is the link labeled ``Cached''.  Clicking this link, we
see the results shown in Figure \ref{fig:google-cache-2005}.  This cached
version has a datestamp of March 3, 2007.  This is typical of SE caches in
that only the most recently cached version is available.

\begin{figure}[!hb]
  \scalebox{0.51}{\includegraphics[clip=false]{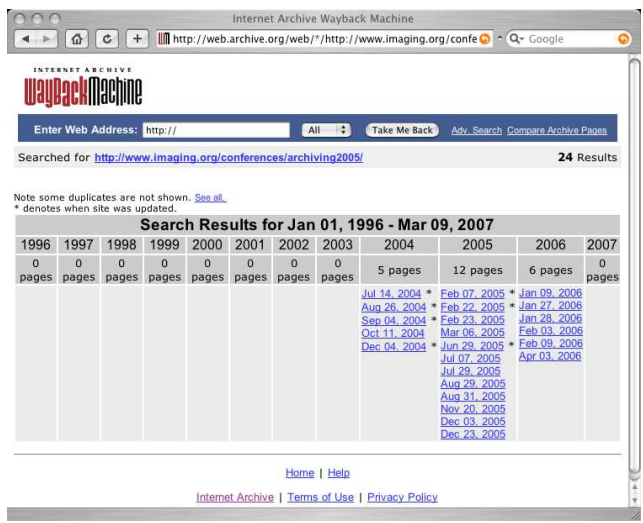}}
  \caption{Archived copies of Archiving 2005 web page.}
  \label{fig:IA-2005}
\end{figure}

In contrast, Figure \ref{fig:IA-2005} shows the multiple datestamped
versions available from the Internet Archive.  The
IA, which first began archiving the Web in 1996 \cite{kahle:preserving},
is unique in its mission of crawling and archiving everything,
with no specific accession policy.  Although the IA is a tremendous
public service, it has some significant limitations.  The first of which
is that the Alexa crawler (Alexa Internet does the crawling for IA)
can be slow to visit a site.  In a previous study, we noted that despite
requesting to be archived, the Alexa crawler never visited our site in
over 100 days (at which point we stopped checking) \cite{smith:observed}.
The second limitation is that even after the site is crawled, the IA will
not make accessible the resources until after 6-12 months have passed \cite{ia:faq}.  In summary,
the IA can be a great boon, but it can be slow to acquire index resources,
and it might not have found them at all.

Besides a study by  Lewandowski et al. \cite{lewandowski:freshness} which examined the freshness of 38 German web pages in SE caches, we are unaware of any research that has characterized the SE caches or attempted to find the overlap of SE caches with the IA.




\section{Methodology}

We chose to study four popular search engines that cache content: Google, MSN,
Yahoo, and Ask.  We used the web search APIs provided by Google,
MSN, and Yahoo for accessing their search results and page scraping
for accessing Ask's search results since they do not provide an API.
Although we have discovered in previous work that the search engine
APIs do not always produce the same result as the web user interface
\cite{McCown07:Agreeing}, we used the APIs because Google
and Yahoo will block access to clients that issue too many queries
\cite{McCown:Google-is-sorry}.\\

In February 2006, we issued 5200 one-term queries (randomly sampled from an English dictionary) to each search engine and randomly chose one of the first 100 results.  We attempted to download the selected URL from the Web and also the cached resource from the SE. We also queried the IA to see how many versions of the URL it had stored for each year, if any. All SE responses, http headers, web pages, cached pages and IA responses were stored for later processing.

Our sampling method produced several biases since it favors pages in English, long and content-rich pages which are more likely to match a query than smaller documents, and those pages that are more popular than others.  New methods \cite{Bar-Yossef:Random} have recently been developed to reduce these biases when sampling from SE indexes and could be used in future experiments.

\section{Cache Content}

We first examine the sampled cache contents and their distribution by top level domain, MIME type and size. We also examine the use of \texttt{noarchive} meta tags and http cache-control directives for keeping content out of SE caches.

\subsection{Cache and Web Overlap}

In Table \ref{tbl:cache-web-overlap} we see the percent of resources from each SE that were cached or not. Within these categories, we break-out those resources that were accessible on the Web or missing (http 4xx or 5xx response or timed-out). Less than 9\% of Ask's indexed contents were cached, but the other three search engines had at least 80\% of their content cached. Over 14\% of Ask's indexed content could not be successfully retrieved from the Web, and since most of these resources were not cached, the utility of Ask's cache is questionable. Google, MSN and Yahoo had far less missing content indexed, and a majority of it was accessible from their cache.

The miss rate column in Table \ref{tbl:cache-web-overlap} is the percent
of time the search engines advertised a link to a cached resource but
returned an error page when the cached resource was accessed.  Ask and
MSN appear to have the most reliable cache access (although Ask's cache
is very small). Note that Google's miss rate is probably higher because
Google's API does not advertise a link to the cached resource; the only
way of knowing if a resource is cached or not is to attempt to access it.

\begin{table}[!h]
\caption{Web and Cache Overlap}
\label{tbl:cache-web-overlap}
\begin{center}
\begin{tabular}{|l|m{0.9cm}m{0.9cm}|m{0.9cm}m{0.9cm}|m{0.8cm}|}
\hline
 & \multicolumn{2}{c}{Cached} & \multicolumn{2}{c}{Not cached} & Miss  \\
 &  Web & Missing  & Web & Missing & rate \\
\hline
Ask    & 8.2\%   & 0.4\% & 77.5\% & 13.9\% & 0.05\% \\
Google & 76.2\%  & 4.3\% & 18.4\%  & 1.1\% & 3.88\% \\
MSN    & 88.7\%  & 4.5\% & 5.0\%  & 1.8\%  & 0.01\% \\
Yahoo  & 76.5\%  & 3.7\% & 17.9\% & 2.0\%  & 1.53\% \\
\hline
\end{tabular}
\end{center}
\end{table}

\begin{figure}[!hb]
\begin{center}
  \scalebox{0.6}{\includegraphics[clip=false]{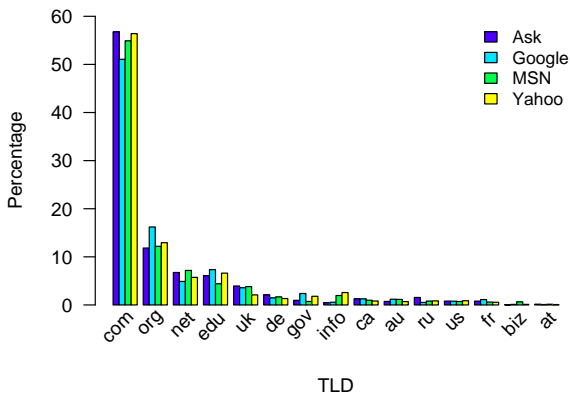}}
  \caption{Distribution of TLDs from sample.}
  \label{fig:tld-distribution}
\end{center}
\end{figure}

\begin{figure*}[!ht]
  \scalebox{0.33}{\includegraphics[clip=true,viewport=0 30 360 250]{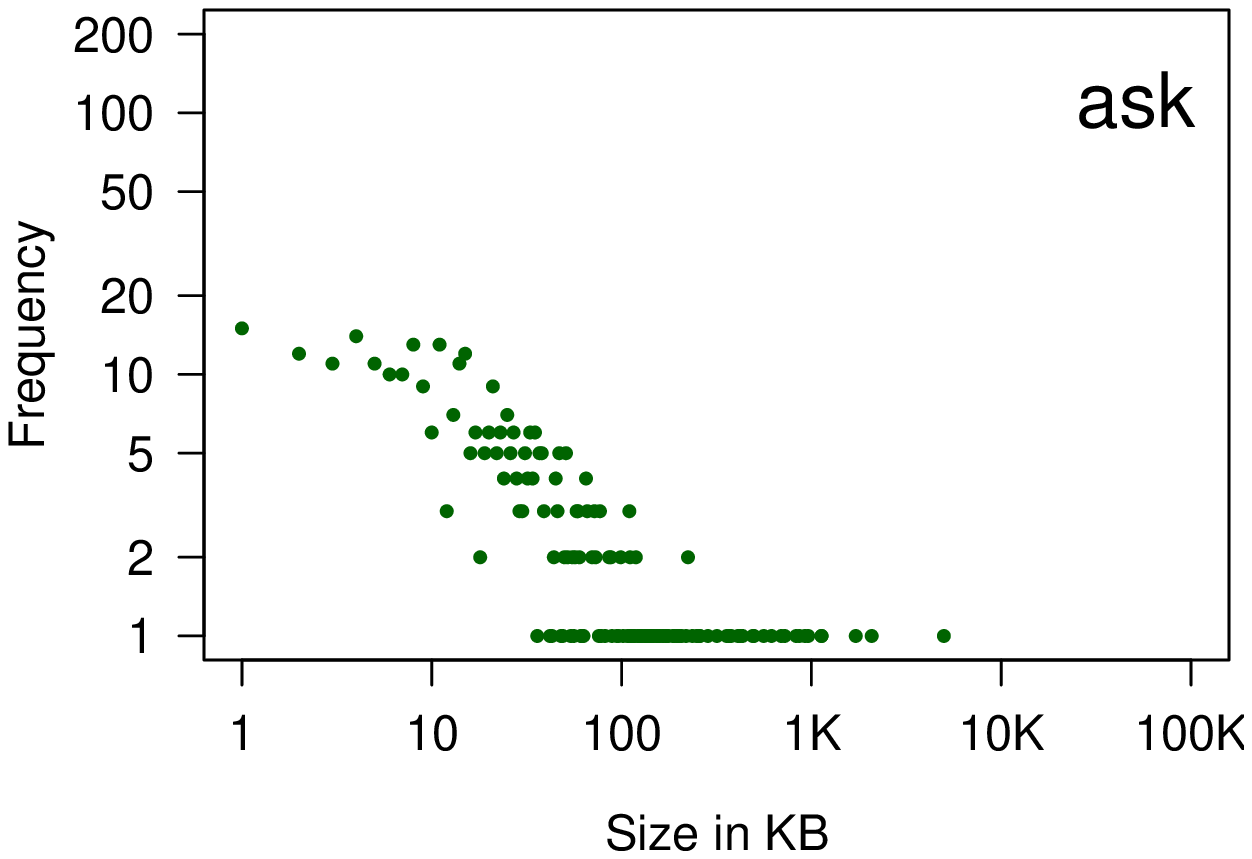}}
  \scalebox{0.33}{\includegraphics[clip=true,viewport=20 30 370 250]{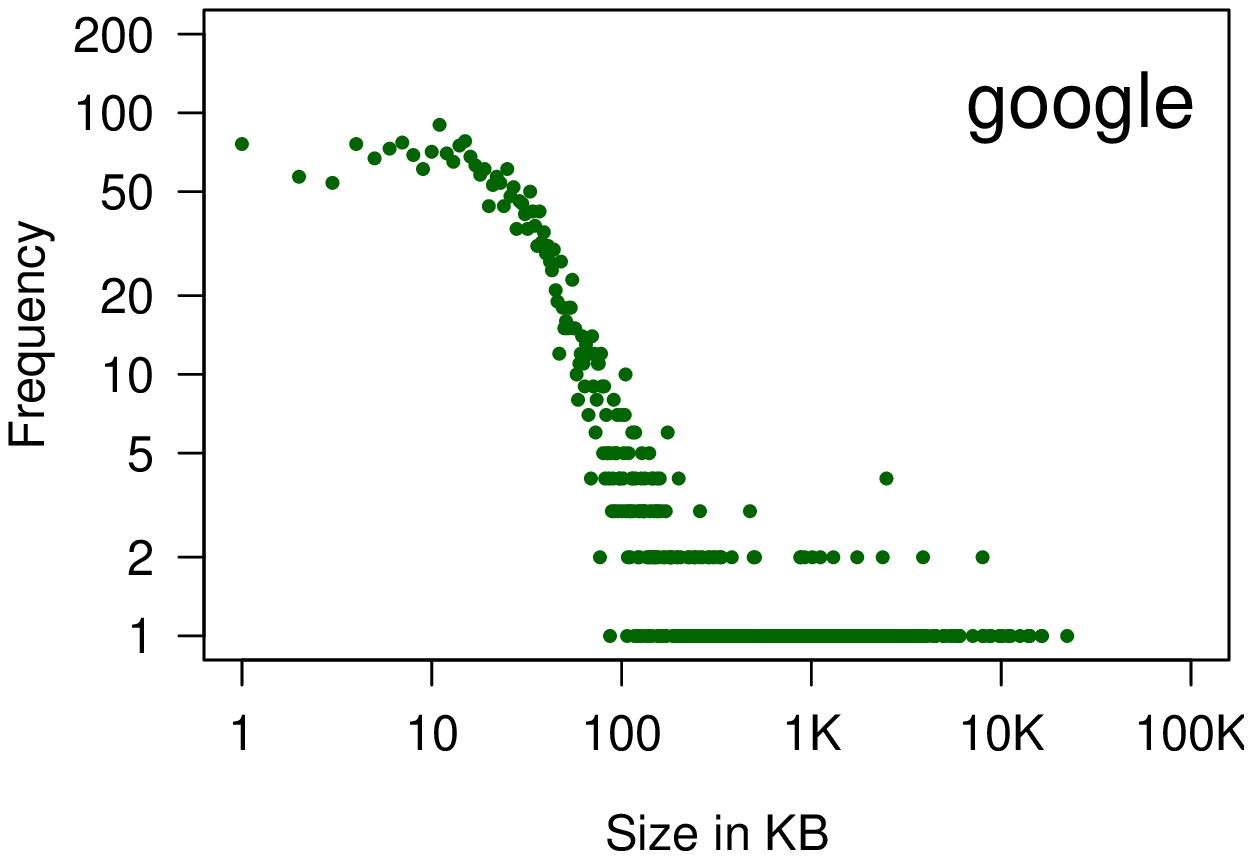}}
  \scalebox{0.33}{\includegraphics[clip=true,viewport=0 30 360 250]{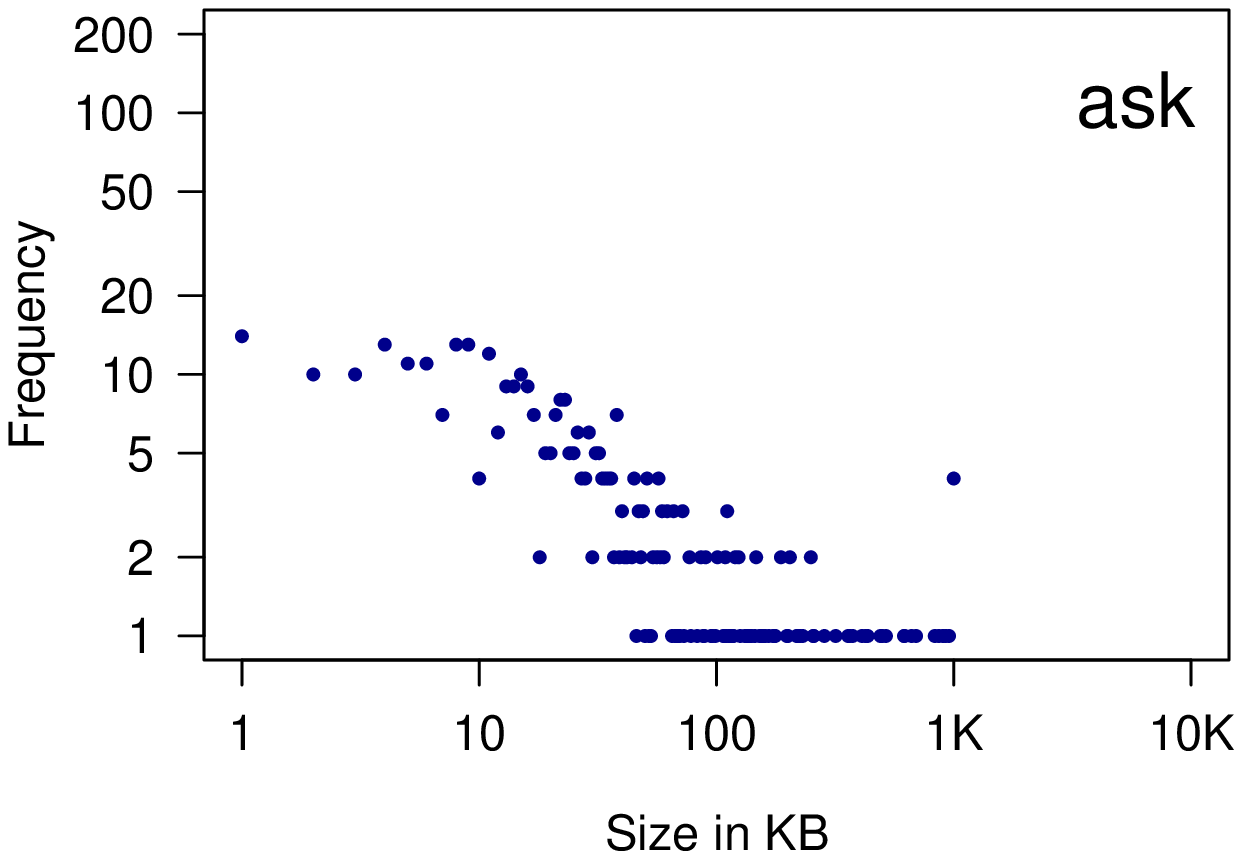}}
  \scalebox{0.33}{\includegraphics[clip=true,viewport=20 30 360 250]{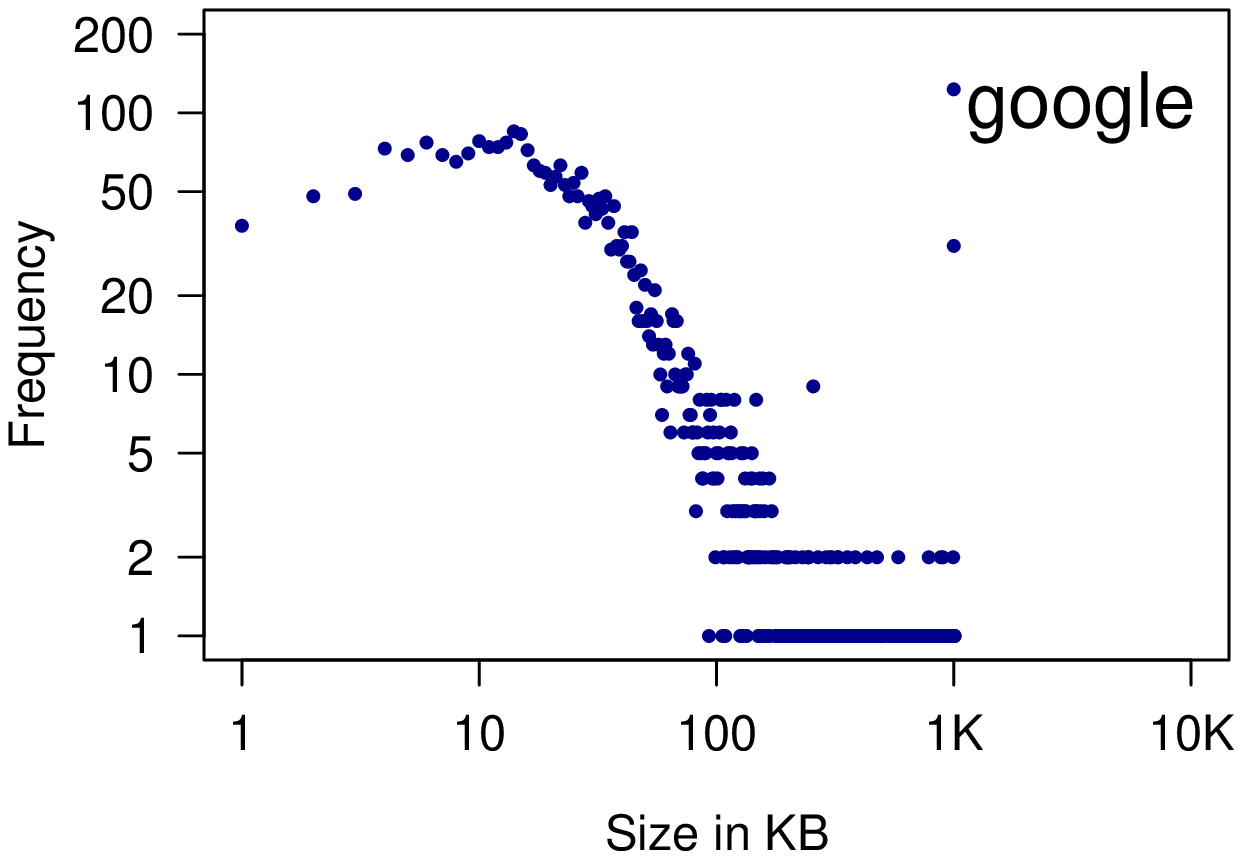}}
  \scalebox{0.33}{\includegraphics[clip=true,viewport=0 0 390 250]{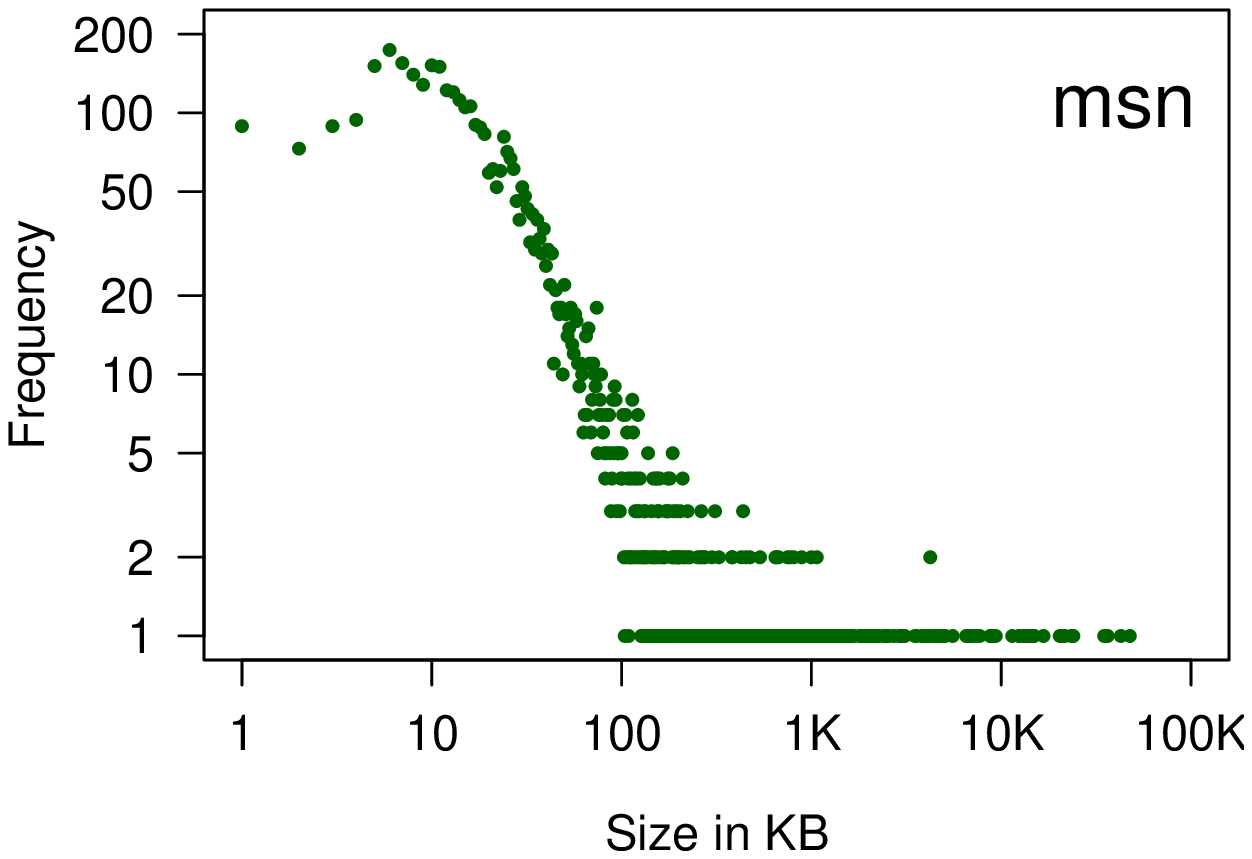}}
  \scalebox{0.33}{\includegraphics[clip=true,viewport=20 0 405 250]{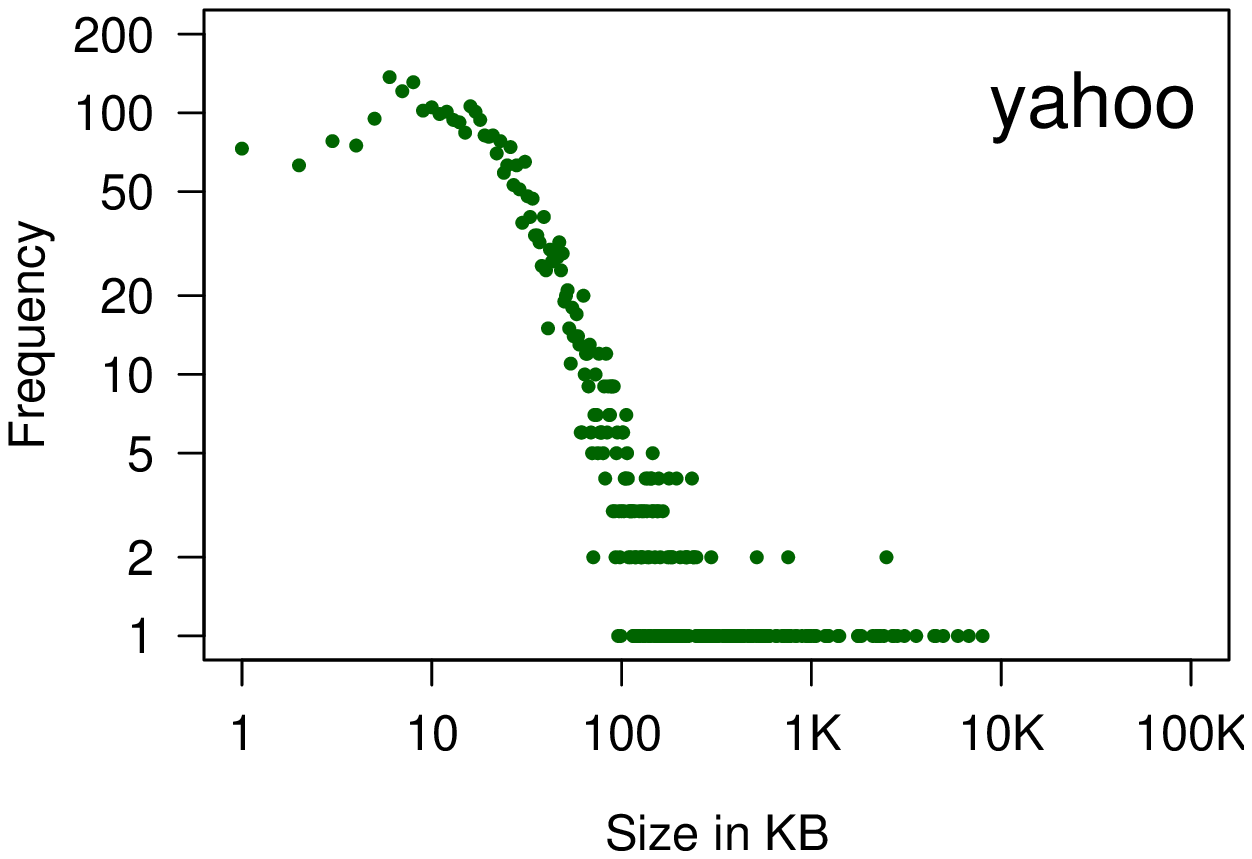}}
  \scalebox{0.33}{\includegraphics[clip=true,viewport=0 0 390 250]{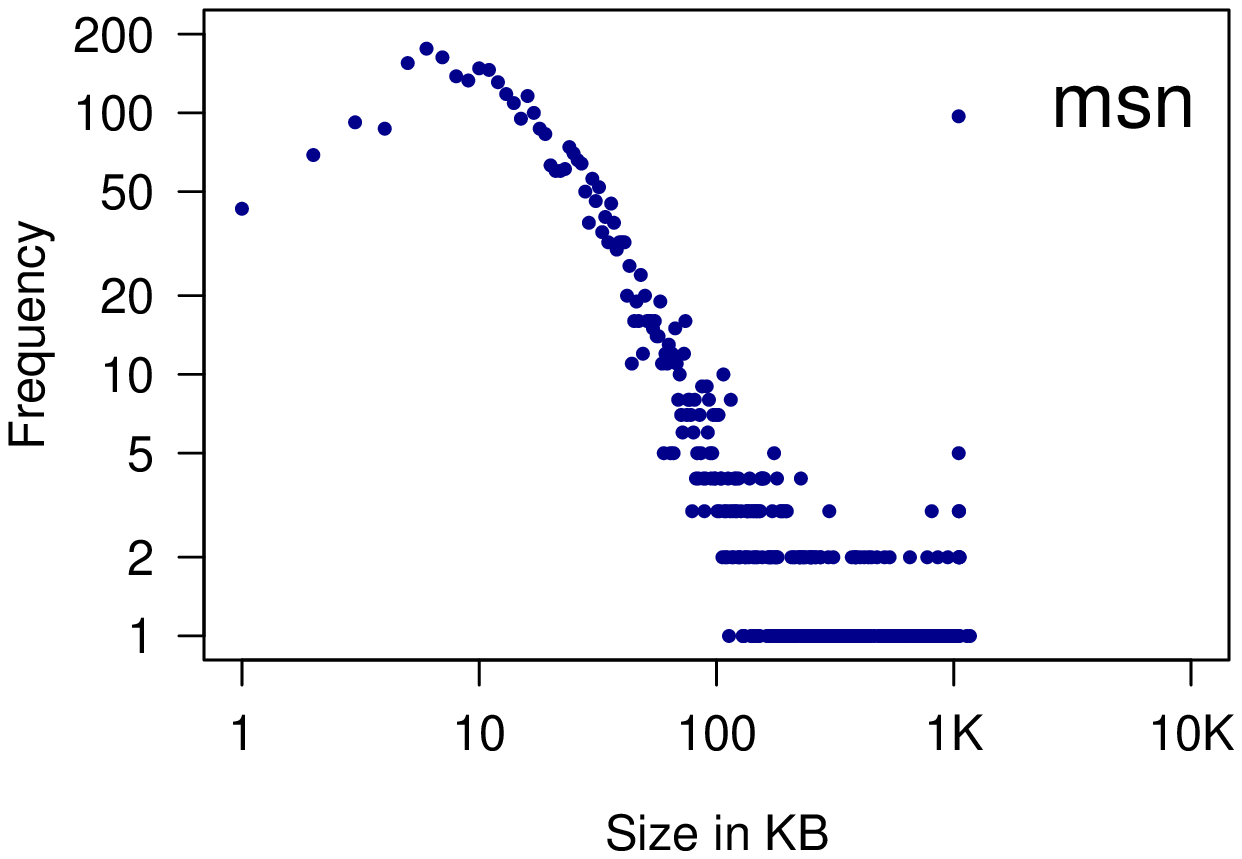}}
  \scalebox{0.33}{\includegraphics[clip=true,viewport=20 0 360 250]{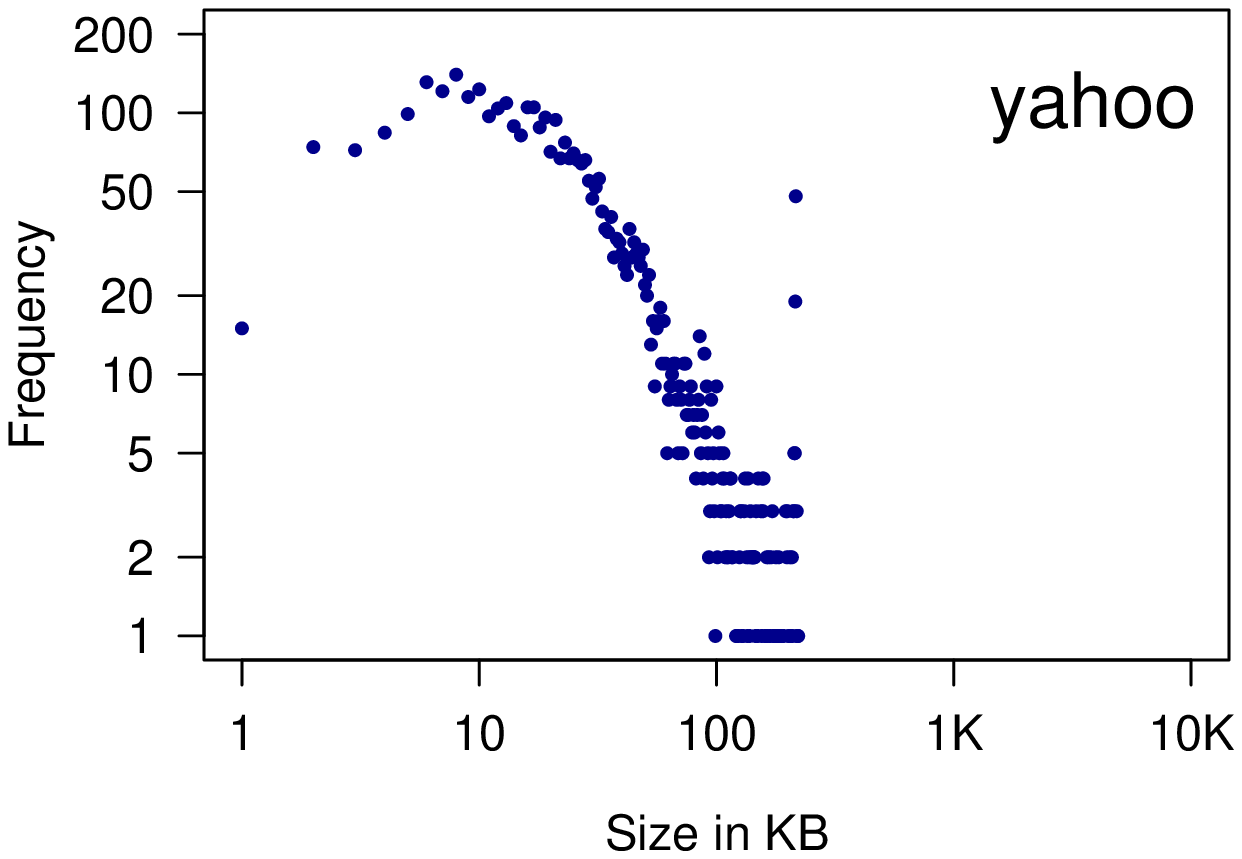}}
  \caption{Distribution of Web file sizes (left) and cached file sizes (right) on log-log scale. Web file size means: Ask = 88 KB, Google = 244 KB, MSN = 204 KB, Yahoo = 61 KB. Cached file size means: Ask = 74 KB, Google = 104 KB, MSN = 79 KB, Yahoo = 34 KB.}
  \label{fig:file-size-distribution}
\end{figure*}

\subsection{Top Level Domain}


Figure \ref{fig:tld-distribution} shows the distribution of the top level
domains (TLDs) of the sampled URLs from each search engine's index (only the top
15 are shown).  Our findings are very similar to the distributions in
\cite{Bar-Yossef:Random}. All four search engines tend to sample equally
from the same TLDs with .com being the largest by far.


\subsection{Content Type}

Table \ref{tbl:indexed-cached-by-type} shows the distribution of resources
sampled from each search engine's index ($Ind$ column). The percent
of those resources that were extracted successfully from cache is given
under the $Cac$ column. HTML was by far the most indexed of all resource
types. Google, MSN and Yahoo provided a relatively high level of access
to all cached resources, but only 10\% of HTML and 11\% of plain text
resources could be extracted from Ask's cache, and no other content type
was found in their cache.











\begin{table}[!h]
\caption{Indexed and Cached Content by Type}
\label{tbl:indexed-cached-by-type}
\begin{footnotesize}
\begin{center}
\begin{tabular}{|l|m{0.39cm}m{0.39cm}|m{0.39cm}m{0.39cm}|m{0.39cm}m{0.39cm}|m{0.39cm}m{0.45cm}|}
\hline
 & \multicolumn{2}{c}{Ask} & \multicolumn{2}{c}{Google} & \multicolumn{2}{c}{MSN} & \multicolumn{2}{c|}{Yahoo} \\
 & Ind & Cac & Ind & Cac & Ind & Cac & Ind & Cac \\
\hline
HTML         & 94\% & 10\% & 88\% & 81\% & 96\% & 95\% & 94\% & 80\% \\
PDF          & 2\%  & 0\%  & 7\%  & 69\% & 3\%  & 89\% & 4\%  & 92\% \\
Plain text   & 4\%  & 11\% & 3\%  & 93\% & 1\%  & 95\% & 1\%  & 96\% \\
MS Office    & 0\%  & 0\%  & 0.7\%  & 76\% & 0.4\%  & 73\% & 0.6\%  & 100\% \\
Other        & 3\%  & 0\%  & 8\%  & 69\% & 3\%  & 89\% & 4\%  & 92\% \\
\hline
\end{tabular}
\end{center}
\end{footnotesize}
\end{table}


We found several media types indexed (but not cached) that we did not expect. We discovered two videos in Google using the Advanced Systems Format (ASF) and an audio file (MPEG) and Flash file indexed by Yahoo. Several XML resource types were also discovered (and some cached): XML Shareable Playlist (Ask), Atom (Google) and RSS feeds (Ask, Google and Yahoo), and OAI-PMH responses (Ask and Google). We did not find any XML types in MSN.

%

\subsection{File Sizes}


In Figure \ref{fig:file-size-distribution} we plot the file size distribution of the live web resources (left) and cached resources (right). The graphs use log-log scale to emphasize the power-law distribution of page sizes which has been observed on the Web \cite{Baeza-Yates:Characterization}. Before calculating the cached resource size, we stripped each resource of the SE header. All four SEs appeared to limit the size of their caches. The limits observed were: Ask: 976 KB, Google: 977 KB, MSN: 1 MB and Yahoo: 215 KB. The caching limits affected approximately 3\% of all resources cached. On average, Google and MSN indexed and cached the largest web resources.

\subsection{Cache Directives}




SEs and the IA use an opt-out policy approach to caching and archiving.  All crawled resources are cached unless a web master uses the Robots Exclusion Protocol (robots.txt) to indicate URL patterns that should not be indexed (which also prevents them from being cached) or if \texttt{noarchive} meta tags are placed in HTML pages. There is currently no mechanism in place to permit a SE to index a non-HTML resource but not cache it.

We found 2\% of the HTML resources from the Web used \texttt{noarchive} meta tags. Only 6\% specifically targeted googlebot, and 96\% targeted all robots (none were targeting the other three SEs). We found only a hand-full of resources with \texttt{noarchive} meta tags that were cached by Google and Yahoo, but it is likely the tags were added after the SE crawlers had downloaded the resources since none of the tags were found in the cached resources.




HTTP 1.1 has a number of cache-control directives that are used to indicate if the requested resource is to be cached, and if so, for how long. Whether or not these directives apply to search engines and web archives is a point of contention \cite{berghel:responsible_web_caching}. One quarter (24\%) of the sampled resources had an http header with Cache-Control set to no-cache, no-store or private, and 62\% of these resources were cached. None of the SEs appeared to respect the cache-control directives since all four SEs cached these resources at the same rate as resources without the header.

\section{Cache Freshness}

We next examine the freshness of the SE caches. A cached copy of a Web resource is fresh if the Web resource has not changed since the last time it was crawled and cached. Once a resource has been modified, the cached resource becomes stale (or ages \cite{cho:effective}).  The staleness of the cache increases until the SE re-crawls the resource and updates its cache.

To measure the staleness the of caches, we examined the Last-Modified
http header of the live resource from the Web and the date from the
cached resource.  Although some servers do not return last modified
dates (typically for dynamically produced resources) or return incorrect
values \cite{clausen04:etag}, it is the best we can do to determine when
the resource was last modified. Not all cached resources contain cache
dates either; Google only reports cached dates for HTML resources, and
Yahoo only reports last modified dates through their API. We calculated
staleness (in days) by subtracting the cached date from the last modified
date. If the cache date was more recent the last modified date, we assigned
a value of 0 to staleness.



\begin{figure}[!hb]
  \scalebox{0.33}{\includegraphics[clip=true,viewport=0 30 360 250]{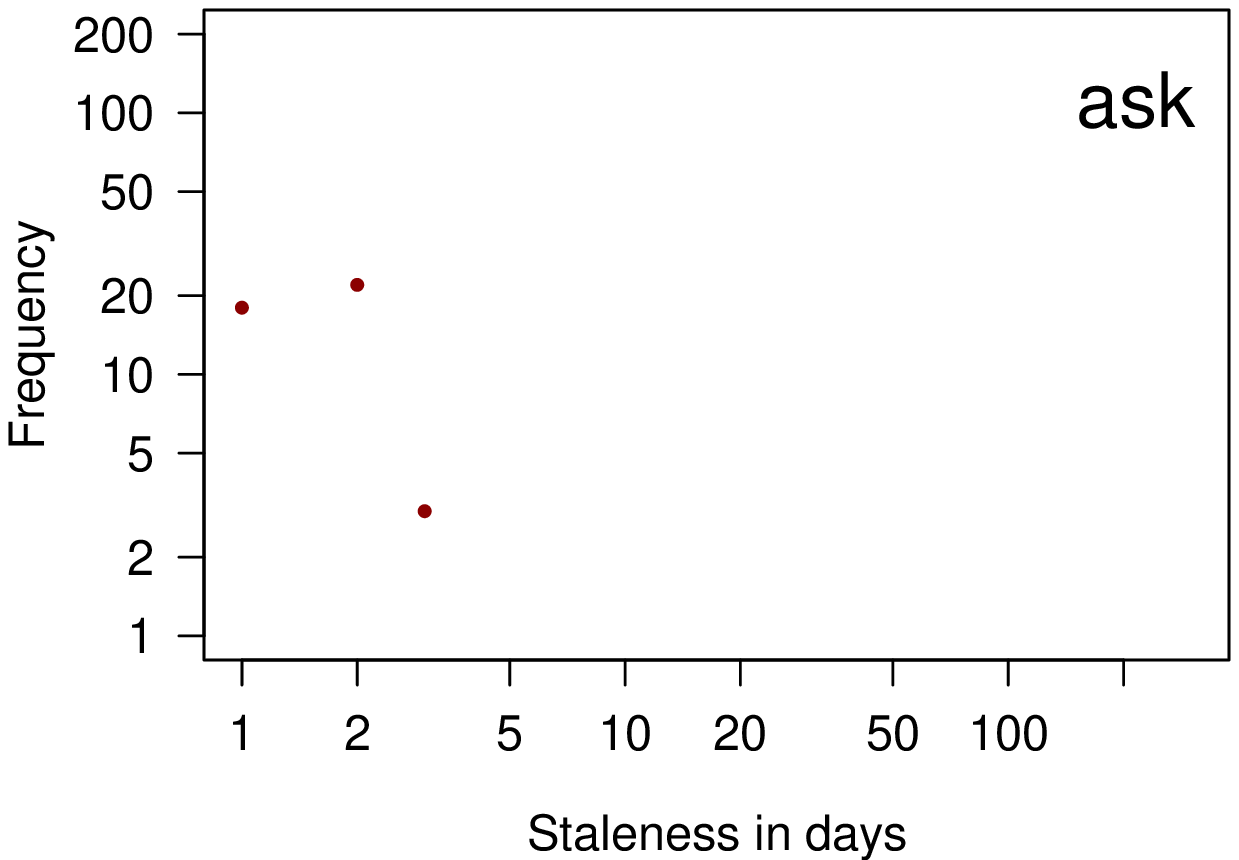}}
  \scalebox{0.33}{\includegraphics[clip=true,viewport=20 30 400 250]{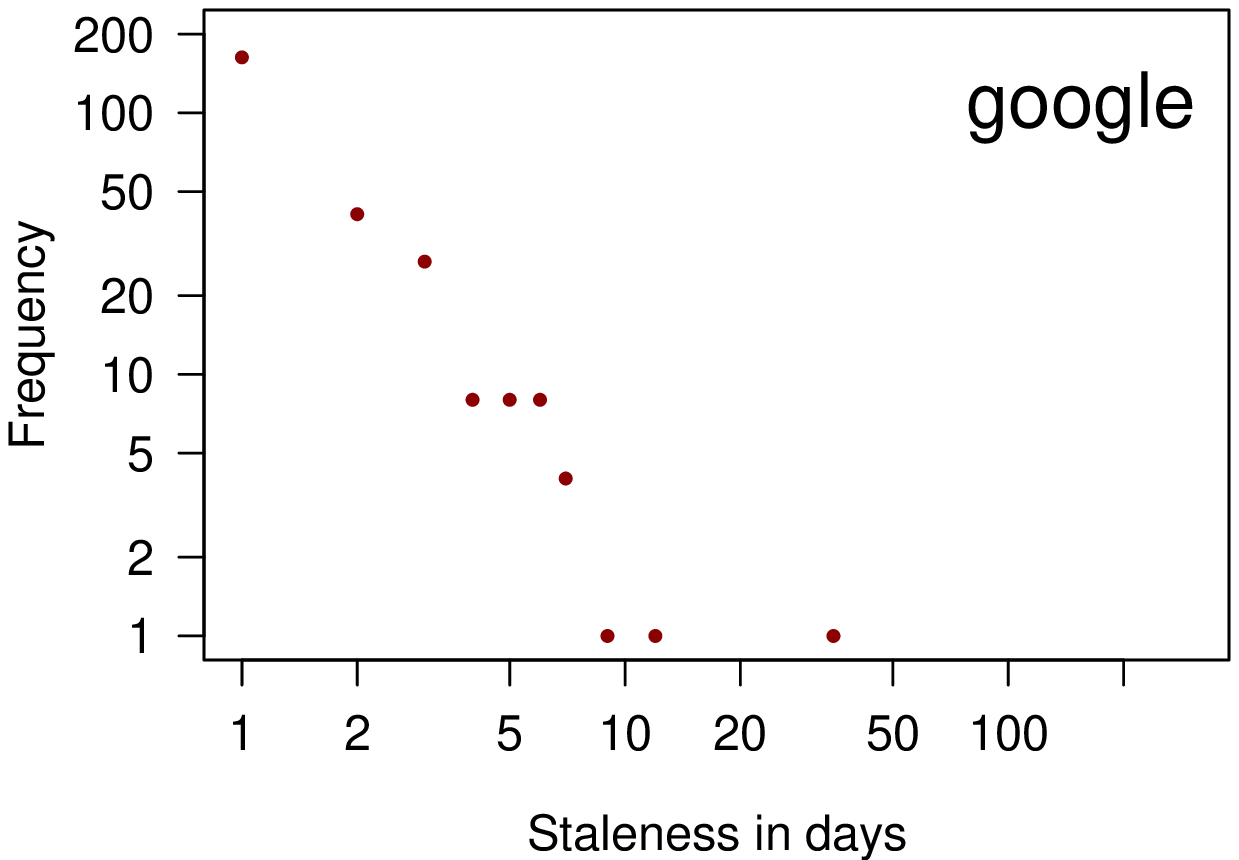}}
  \scalebox{0.33}{\includegraphics[clip=false]{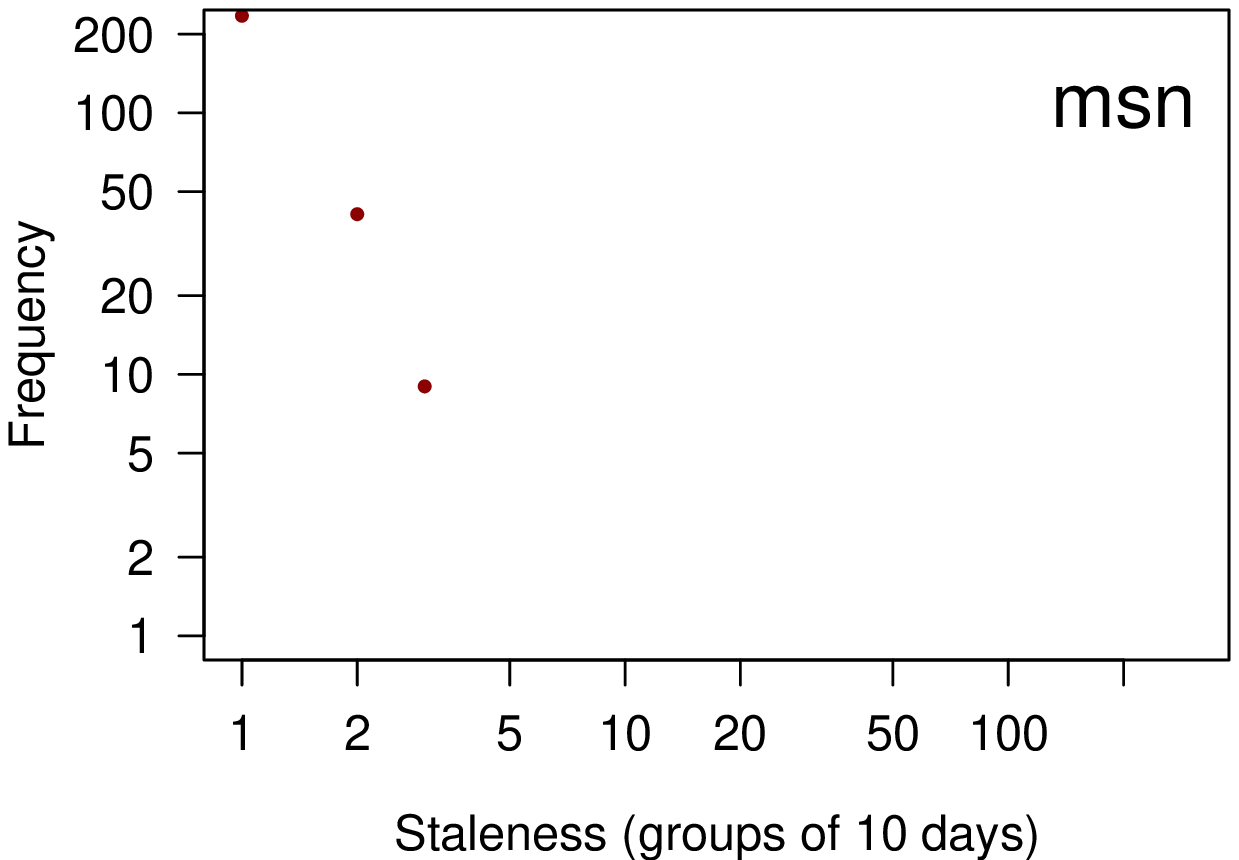}}
  \scalebox{0.33}{\includegraphics[clip=true,viewport=20 0 400 250]{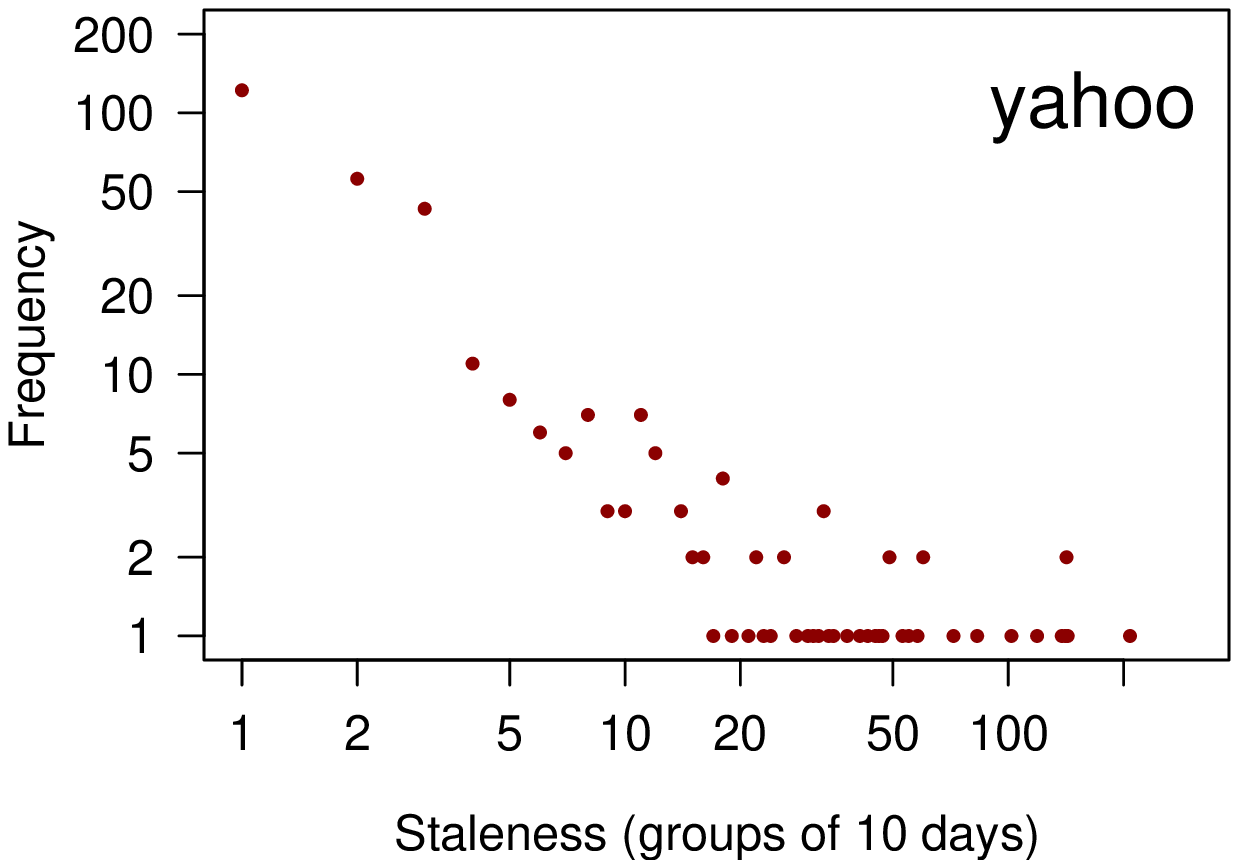}}
  \caption{Distribution of staleness on log-log scale.}
  \label{fig:staleness-distribution}
\end{figure}

Only 46\% of the live pages had a valid http Last-Modified timestamp,
and of these, 71\% also had a cached date. We found 84\% of the resources
were up-to-date.  The descriptive statistics for resources that were
at least one day stale are given in Table \ref{tbl:staleness}, and the distribution is shown in
Figure \ref{fig:staleness-distribution}. Although
Google had the largest amount of stale cached pages, Yahoo's pages were
on average more stale. MSN had the fewest amount of stale pages and
nearly the most up-to-date set of pages.



\begin{table}[!h]
\caption{Staleness of Search Engine Caches (in Days)}
\label{tbl:staleness}
\begin{center}
\begin{tabular}{|l|m{1.0cm}m{0.7cm}m{0.7cm}m{0.7cm}m{0.5cm}m{0.5cm}|}
\hline
 & \% Stale & Mean & Median  & Std & Min & Max  \\
\hline
Ask    & 16\% & 7.5  & 12.0 & 5.5 & 2 & 25 \\
Google & 20\% & 15.1 & 7.0 & 26.2 & 1 & 343 \\
MSN    & 12\% & 6.7  & 5.0 & 5.2 & 1 & 27 \\
Yahoo  & 17\% & 92.1 & 17.0 & 220.9 & 1 & 1422 \\
\hline
\end{tabular}
\end{center}
\end{table}

We also wanted to know how similar the cached resources were compared to the live resources from the Web.  We would expect up-to-date cached resources to be identical or nearly identical to their Web counter-parts. We would also expect web resources in formats that get converted into HTML (e.g., PDF, PostScript and Microsoft Office) to be very similar to their cached counterparts in terms of word order.  When comparing live resources to crawled resources, we counted the number of shared shingles (of size 10) between the two documents after stripping out all HTML (if present). Shingling \cite{broder:syntactic} is a popular technique for quantifying similarity of text documents when word-order is important.



We found that 19\% of the cached resources were identical to their live counterparts, 21\% if examining just HTML resources. On average, resources shared 72\% of their shingles. This implies that although most web resources are not replicated in caches byte-for-byte, most of them are very similar to what is cached.

\begin{figure}[!hb]
  \scalebox{0.33}{\includegraphics[clip=true,viewport=0 30 360 250]{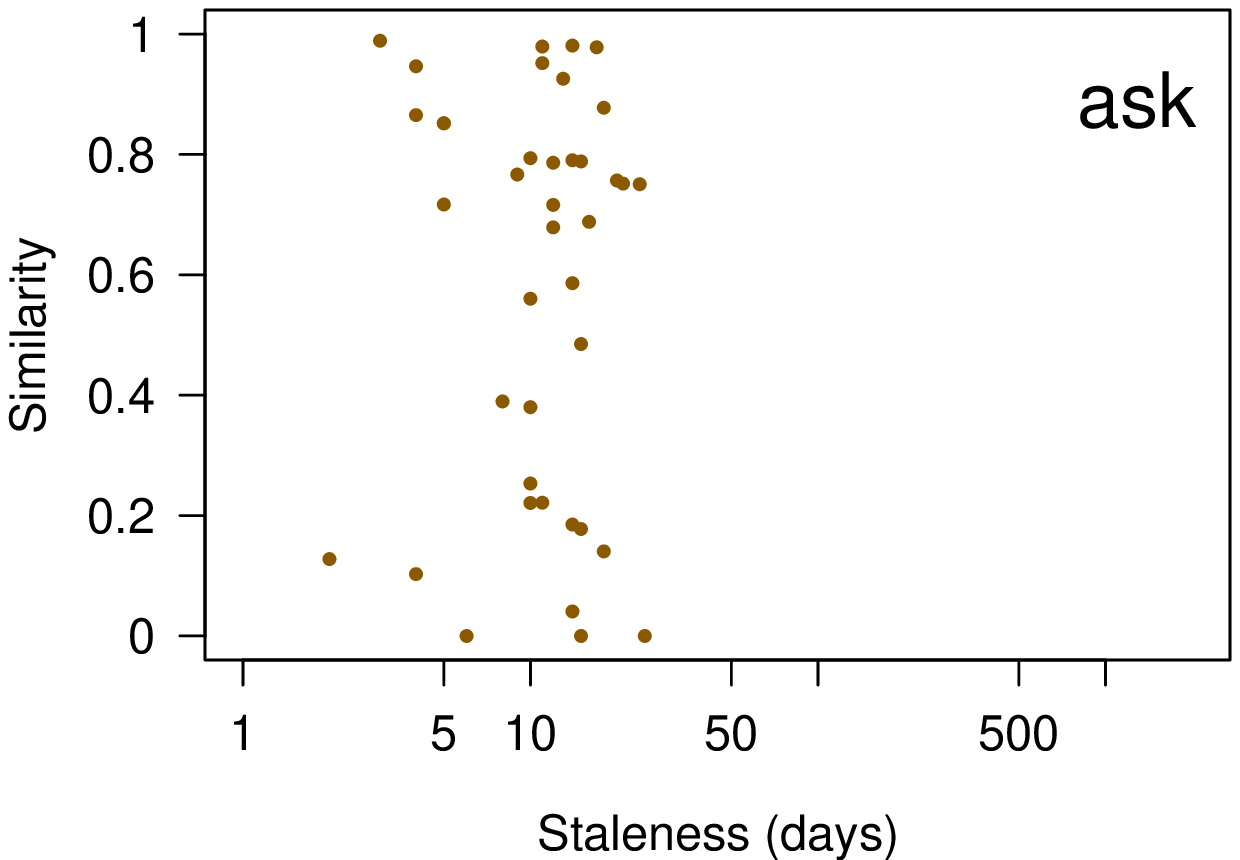}}
  \scalebox{0.33}{\includegraphics[clip=true,viewport=20 30 400 250]{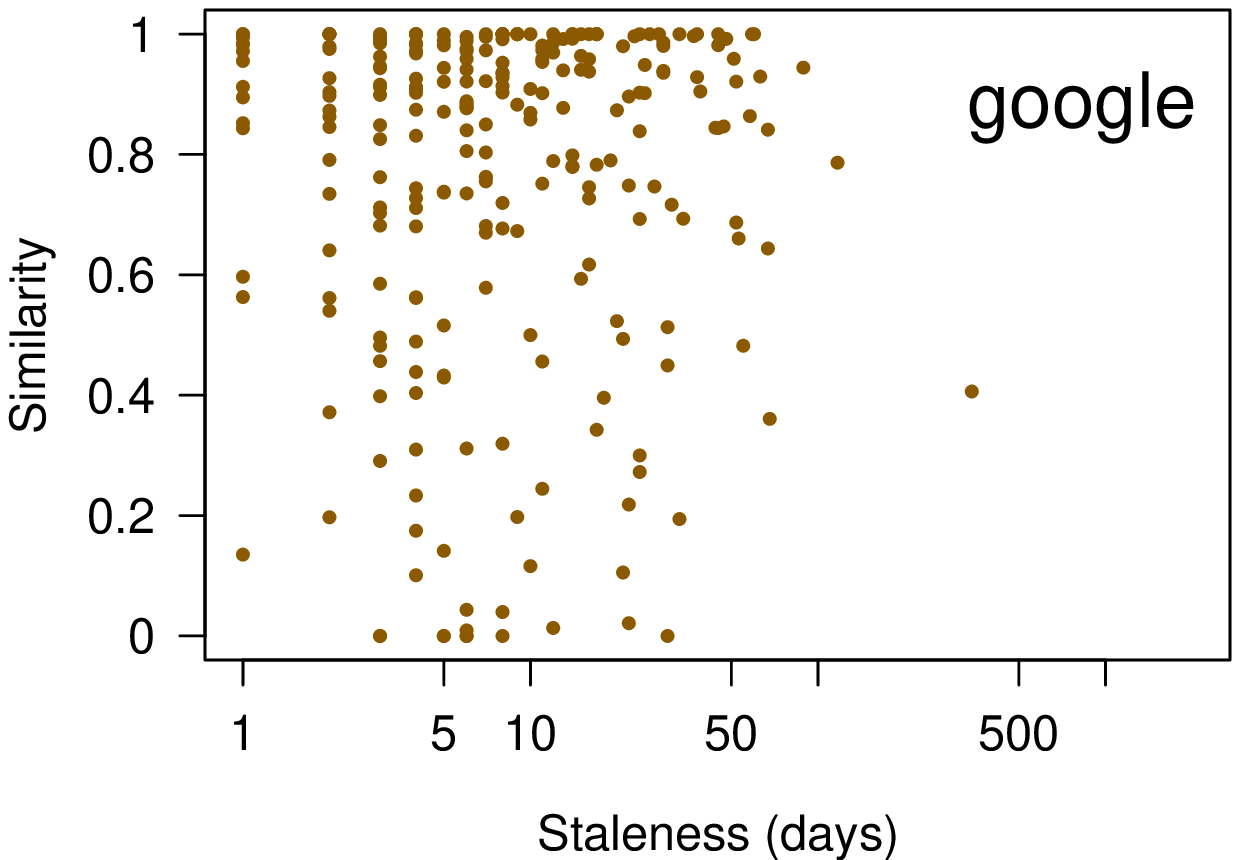}}
  \scalebox{0.33}{\includegraphics[clip=false]{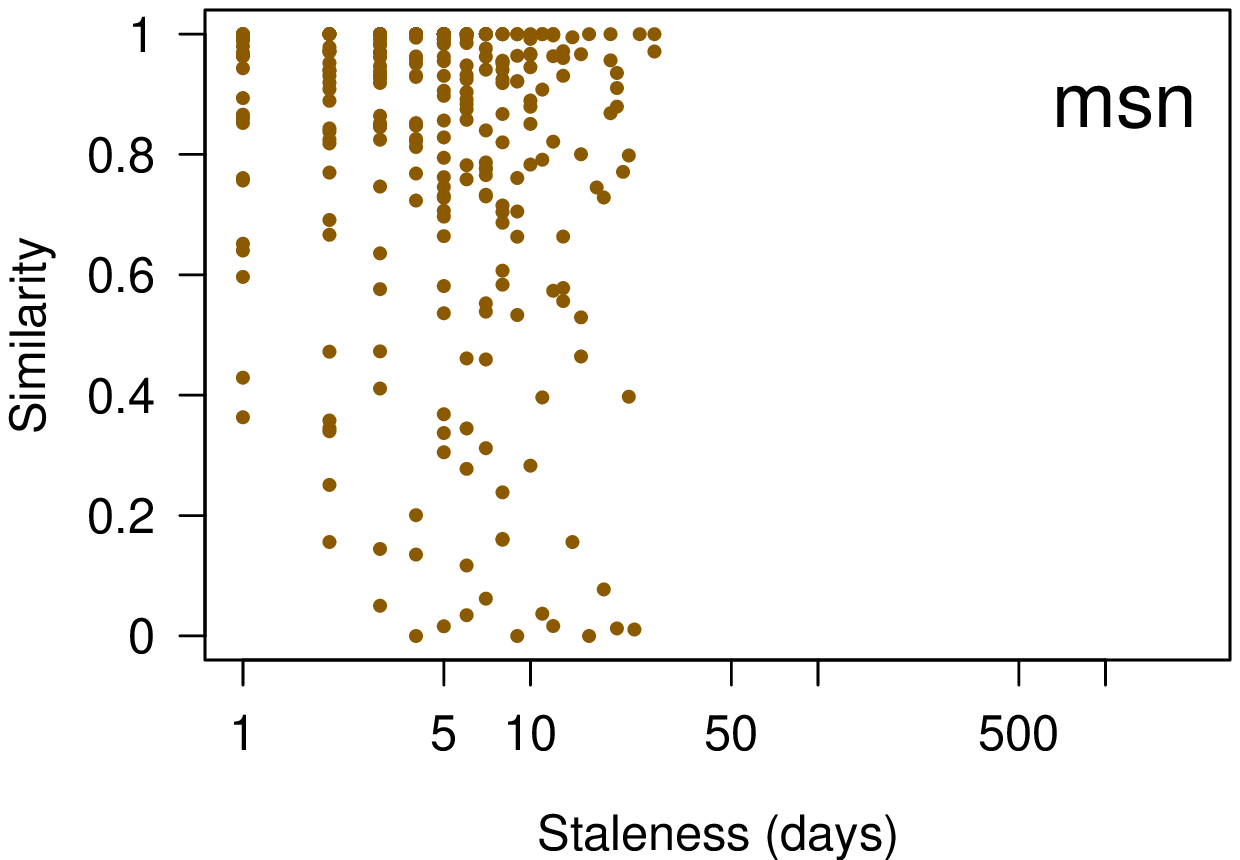}}
  \scalebox{0.33}{\includegraphics[clip=true,viewport=20 0 400 250]{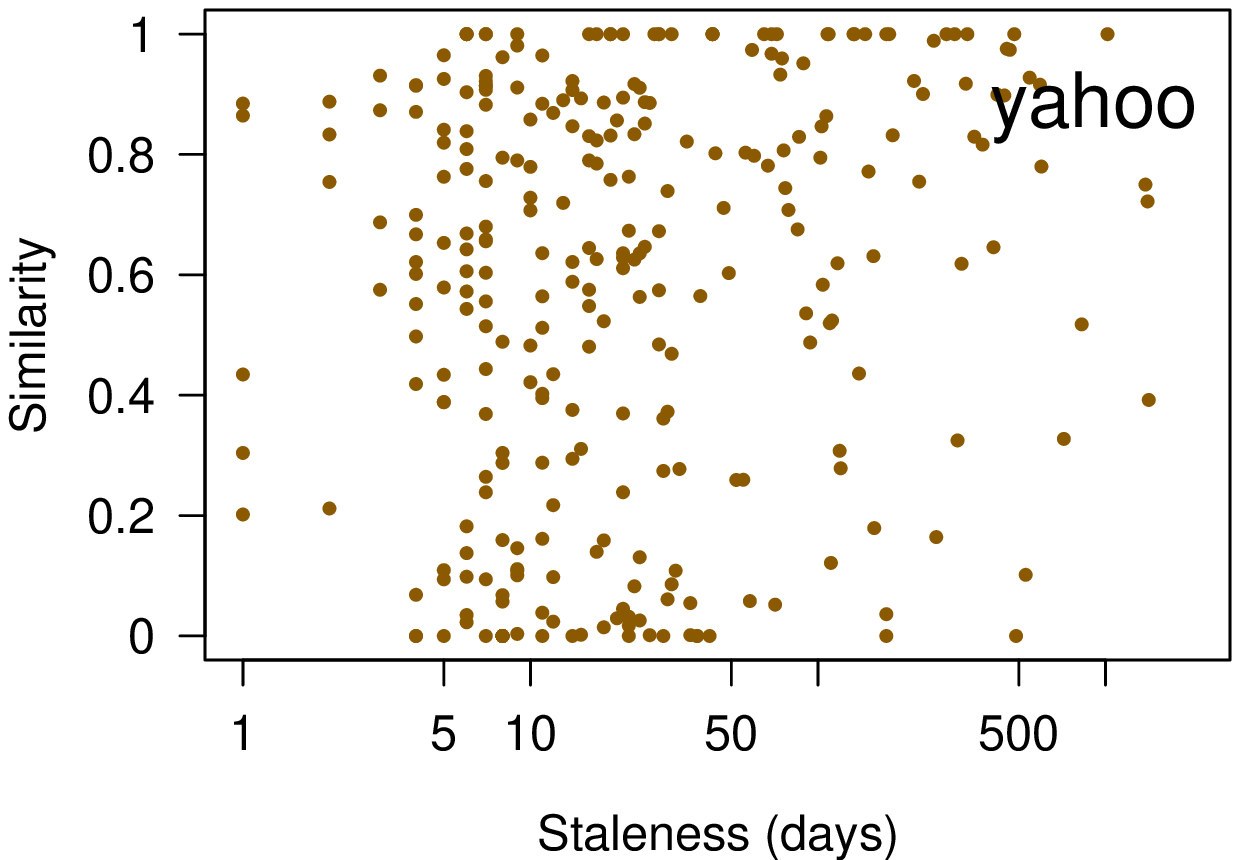}}
  \caption{Scatterplots of similarity vs. staleness (x-axis is on a log scale).}
  \label{fig:shingles-vs-staleness}
\end{figure}

In Figure \ref{fig:shingles-vs-staleness} we have plotted each resource's `similarity' value (percent of shared shingles) vs. its staleness.  The busy scatterplots indicate there is no clear relationship between similarity and staleness; a cached resource is likely to be just as similar as its live Web counterpart if it is one or 100 days stale.


\section{Overlap with Internet Archive}

\begin{figure}
  \scalebox{0.83}{\includegraphics[clip=true,viewport=0 0 300 110]{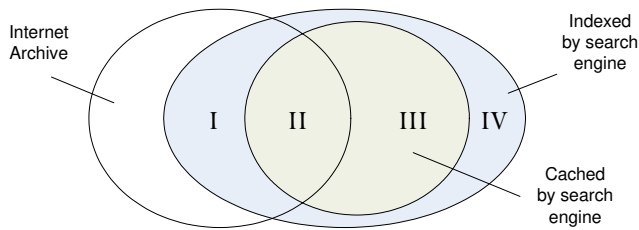}}
  \caption{Venn diagram showing overlap of SE caches with IA.}
  \label{fig:ia-overlap}
\end{figure}

We were interested in knowing how much content indexed and cached by the four SEs were also archived by the Internet Archive. Figure \ref{fig:ia-overlap} shows a Venn diagram illustrating how some resources held by the IA are indexed by SEs (I) and/or cached (II). But there are some indexed (IV) and cached (III) resources that are not available in the IA.

Table \ref{tbl:ia-overlap} shows the overlap of sampled URLs within IA. MSN had the largest overlap with IA (52\%) and Yahoo the smallest (41\%). On average, only 46\% of the sampled URLs from all four SEs were available in IA.

\begin{table}[!h]
\caption{Internet Archive Overlap}
\label{tbl:ia-overlap}
\begin{center}
\begin{tabular}{|l|m{1.2cm}m{1.3cm}|m{1.2cm}m{1.3cm}|}
\hline
 & \multicolumn{2}{c}{In IA} & \multicolumn{2}{c|}{Not in IA}  \\
SE &  Cached (II) & No cache (I)  & Cached (III) & No cache (IV)  \\
\hline
Ask       & 9.2\%  & 36.0\%  & 0.3\%   & 54.5\% \\
Google    & 40.7\% &  3.7\%  & 50.3\%  & 5.3\% \\
MSN       & 51.1\% &  1.1\%  & 43.7\%  & 4.1\% \\
Yahoo     & 39.3\% &  1.8\%  & 47.7\%  & 11.2\% \\
\hline
\end{tabular}
\end{center}
\end{table}



\begin{figure}
\begin{center}
  \scalebox{0.6}{\includegraphics[clip=false]{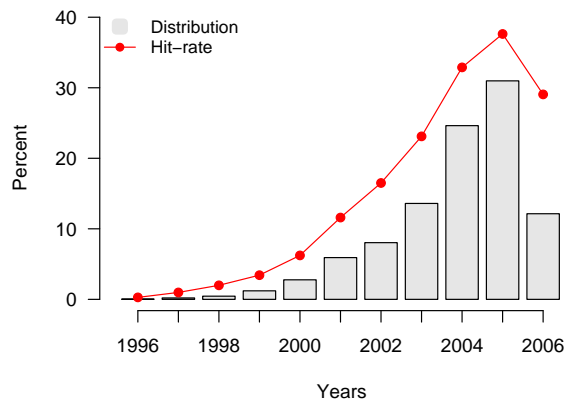}}
  \caption{Distribution and hit rate of sampled URLs with IA.}
  \label{fig:ia-dist}
\end{center}
\end{figure}

In Figure \ref{fig:ia-dist} we have plotted the distribution of the archived resources which shows an almost exponential increase each year. We suspect there were far fewer resources in 2006 since the IA is 6-12 months out of date. The hit-rate line in Figure \ref{fig:ia-dist} is the percent of time the IA had at least one resource archived for that year. It is interesting to note that although the number of resources archived in 2006 was half that of 2004, the hit rate of 29\% almost matched 2004's 33\% hit rate.


\section{Conclusions}

In this study, we have characterized the caches of Ask, Google, MSN
and Yahoo by randomly choosing results from the top
100 hits based on dictionary-based queries.  From a
digital preservation perspective, Ask was of limited utility; it had the
fewest resources cached (9\%), and although 14\% of the resources it had indexed were unavailable from the Web, only 3\% of them were accessible from their cache.  The resources from Google (80\%), MSN (93\%) and Yahoo (80\%) were
cached much more frequently, and all had limited cache miss rates.
Top level domains appear to be represented in all four SE caches with
roughly the same distribution.  We found \texttt{noarchive} meta tags
were infrequently used (2\%) in sampled HTML resources, and SEs did not appear to
respect http cache-control headers, two advantages from a preservation perspective.

Search engines primarily index HTML, but of the resources that are indexed, all
SEs but Ask cached non-HTML resources with about the same frequency.  All SEs seemed to have an upper bound on cached resources of about 1 MB
except for Yahoo which appears to have an upper bound of 215 KB; this only affected 3\% of all cached resources.  The ``staleness''
of the cached resources ranged from 12\% (MSN) to 20\% (Google), and median
staleness ranged from 5 days (MSN) to 17 days (Yahoo).

While the IA provides a preservation service for public web pages, its well-known
limitations of crawling frequency and 6-12 month delay in processing crawled
resources limits its effectiveness.  We found that the IA contained
only 46\% of the resources available in SE caches.  More importantly, the number
of resources available in neither a SE cache nor the IA is quite low: 4\% for MSN,
5\% for Google and 11\% for Yahoo.  Again, Ask (55\%) performs poorly.

Search engines provide access to cached copies as a secondary service
to guard against temporary unavailability of the indexed resources.
But given enough SEs and their respective scale, it becomes
possible (especially in combination with the IA) to build sophisticated
digital preservation services using SE caches.

\section{Acknowledgments}

We would like to thank Giridhar Nandigam (ODU) for writing the search engine querying software.
This work is supported in part by NSF Grant IIS-0610841.




\small


\begin{biography}

Frank McCown is a Ph.D. candidate in the Computer Science Department of Old Dominion University. He has been an instructor at Harding University since 1997. His research interests include web crawling, digital preservation, search engines and digital libraries. Michael L. Nelson joined the Computer Science Department at Old Dominion University in 2002. He worked at NASA Langley Research Center from 1991-2002. His research interests include repository-object interaction and digital preservation.


\end{biography}

\end{document}